\documentclass[journal,twoside,web]{ieeecolor}

\usepackage{jsen}
\usepackage{cite}
\usepackage{amsmath,amssymb,amsfonts}
\usepackage{algorithmic}
\usepackage{graphicx}
\usepackage{textcomp}
\usepackage{wrapfig}

\newcommand{\papername}{DCAMamba}

\def\BibTeX{{\rm B\kern-.05em{\sc i\kern-.025em b}\kern-.08em
    T\kern-.1667em\lower.7ex\hbox{E}\kern-.125emX}}
\definecolor{abstractbg}{rgb}{0.89804,0.94510,0.83137}
\begin{document}
\title{\papername: Mamba-based Rapid Response DC Arc Fault Detection }

\author{Lukun Wang, Ruxue Zhao, Wancheng Feng, Pu Sun, and Chunpeng Tian
\thanks{Lukun Wang, He is 
now with the College of intelligent Equipment, Shandong University of Science and Technology, Taian, China.  (e-mail: wanglukun@sdust.edu.cn).}
\thanks{Ruxue Zhao, She is 
now with the College of intelligent Equipment, Shandong University of Science and Technology, Taian, China.  (e-mail: zhaoruxue@sdust.edu.cn).}
\thanks{Wancheng Feng, He is 
now with the College of intelligent Equipment, Shandong University of Science and Technology, Taian, China.  (e-mail: fengwancheng@sdust.edu.cn).}
\thanks{Pu Sun, He was with the College of intelligent Equipment, Shandong University of Science and Technology, Taian, China. 
He is now in Jinan Tobacco Monopoly Bureau, Jinan, China.  (e-mail:sunpu@sdust.edu.cn ).}
\thanks{Chunpeng Tian, He is 
now with the College of intelligent Equipment, Shandong University of Science and Technology, Taian, China.  (e-mail: tianchunpeng@sdust.edu.cn).}}

\IEEEtitleabstractindextext{%
\fcolorbox{abstractbg}{abstractbg}{%
\begin{minipage}{\textwidth}%
\begin{wrapfigure}[15]{r}{3.2in}%
\includegraphics[width=3in]{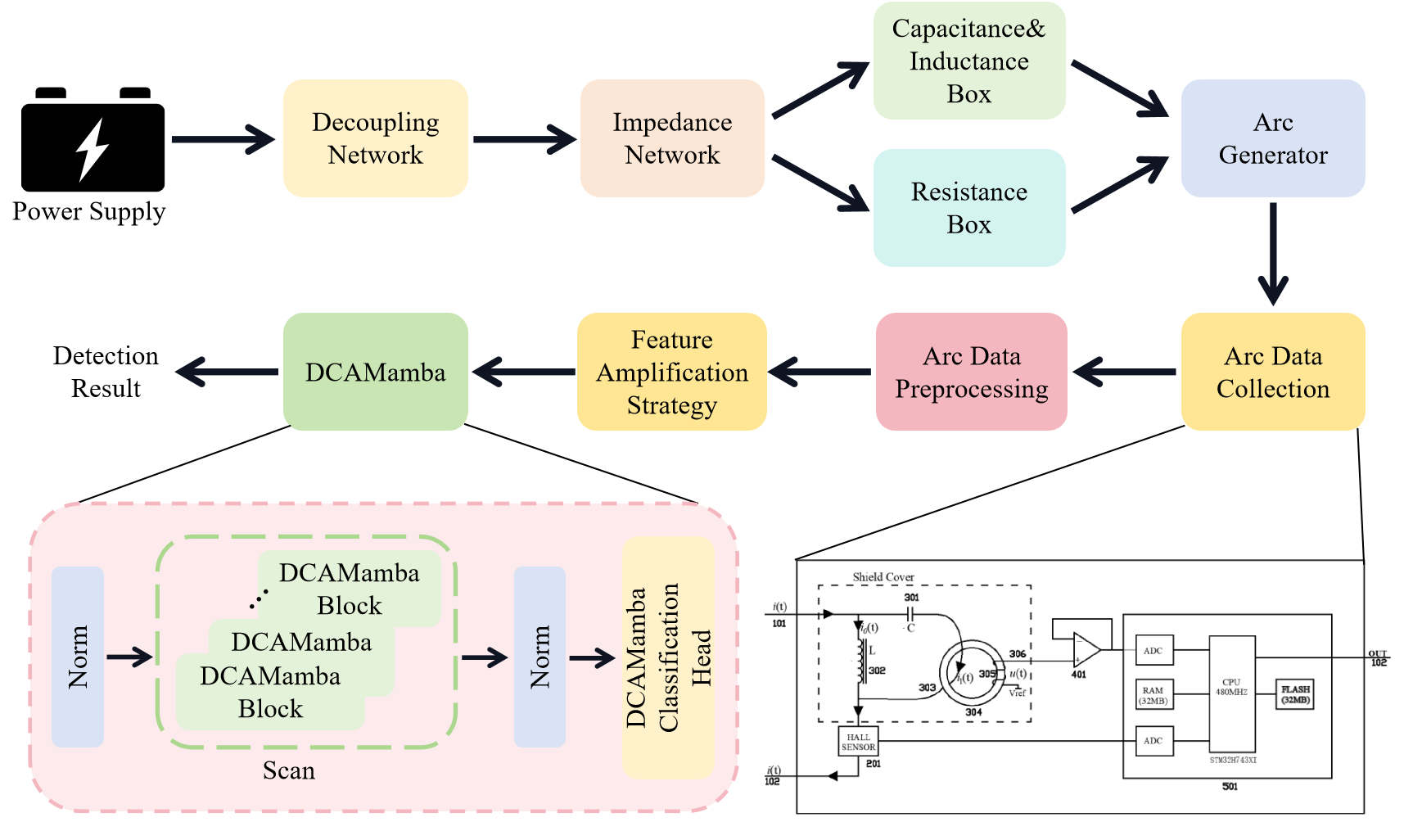}%
\end{wrapfigure}%
\begin{abstract}
In electrical equipment, even minor contact issues can lead to arc faults. Traditional methods often struggle to balance the accuracy and rapid response required for effective arc fault detection. To address this challenge, we introduce DCAMamba, a novel framework for arc fault detection. Specifically, DCAMamba is built upon a state-space model (SSM) and utilizes a hardware-aware parallel algorithm, designed in a cyclic mode using the Mamba architecture. To meet the dual demands of high accuracy and fast response in arc fault detection, we have refined the original Mamba model and incorporated a Feature Amplification Strategy (FAS), a simple yet effective method that enhances the model's ability to interpret arc fault data. Experimental results show that DCAMamba, with FAS, achieves a 12$\%$ improvement in accuracy over the original Mamba, while maintaining an inference time of only 1.87 milliseconds. These results highlight the significant potential of DCAMamba as a future backbone for signal processing. Our code will be made open-source after peer review.

\end{abstract}

\end{minipage}}}

\maketitle
\thispagestyle{empty}

\section{Introduction}
\label{sec:introduction}
With the global demand for sustainable energy continuously increasing, photovoltaic (PV) power generation, as a clean and renewable energy source, has rapidly developed. By 2024, the global cumulative installed capacity of PV systems reached 1.6 TW, with 407 GW to 446 GW of new installations added in 2023 alone. The scale of the PV industry continues to grow. As the systems age, electrical equipment degradation, inverter malfunctions, and component failures in PV systems may lead to arc faults, which can trigger fires, causing severe economic losses, environmental damage, and safety risks.

\begin{figure}[t]
    \centering
    \includegraphics[width=8.5cm]{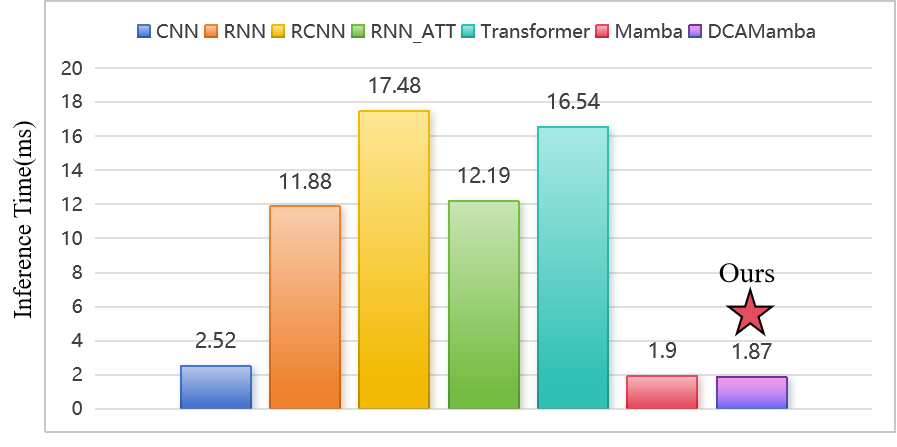}
    \caption{
        The Inference Time of our method takes only 1.87$ms$. In the same experimental environment, compared with the current mainstream methods, DCAMamba is even faster than simple CNN.
    }
    \label{fig:time}
\end{figure}

In electrical circuits, when high voltage breaks down a gaseous medium (such as air), the gas molecules are ionized, forming a conductive channel. The current passing through this channel results in discharge, known as an arc. During arc formation, high temperatures, intense light radiation, and electromagnetic interference are produced, which significantly increase the risk of fire. To mitigate the dangers posed by arc faults, it is crucial to detect them in a timely manner.

Arc faults can be classified into three types~\cite{thakur2023advancements}: series arc faults, parallel arc faults, and ground arc faults~\cite{artale2017arc}. Series arc faults are typically caused by poor contact or insulation damage, occurring in the series portion of the circuit. Parallel arc faults are usually due to wiring errors or equipment failure, occurring in the parallel branches of the circuit. Ground arc faults occur in the grounding system when a part of the circuit comes into contact with the ground, forming an arc through the grounding loop. Each type of arc fault has distinct characteristics, leading to different fault detection approaches.

In photovoltaic (PV) systems, DC arc faults are more dangerous and harder to eliminate than AC arc faults because DC current does not have a natural zero-crossing point. Once an arc is formed, it is difficult to extinguish automatically, which increases the complexity of detection and control. Series arc faults are more challenging to detect compared to parallel arc faults. Parallel circuits have higher total power, and when arc faults occur, circuit breakers can quickly disconnect the power to prevent damage. However, series circuits in PV systems typically have more connections and lower currents, which are insufficient to trigger circuit breakers. Furthermore, PV systems contain thousands of connection points, and faults can occur at any location, such as junctions, connections, or during transmission, which further complicates the detection of series arc faults~\cite{qi2022cybertwin,qi2023adaptive}.

\begin{figure}[t]
    \centering
    \includegraphics[width=8.5cm]{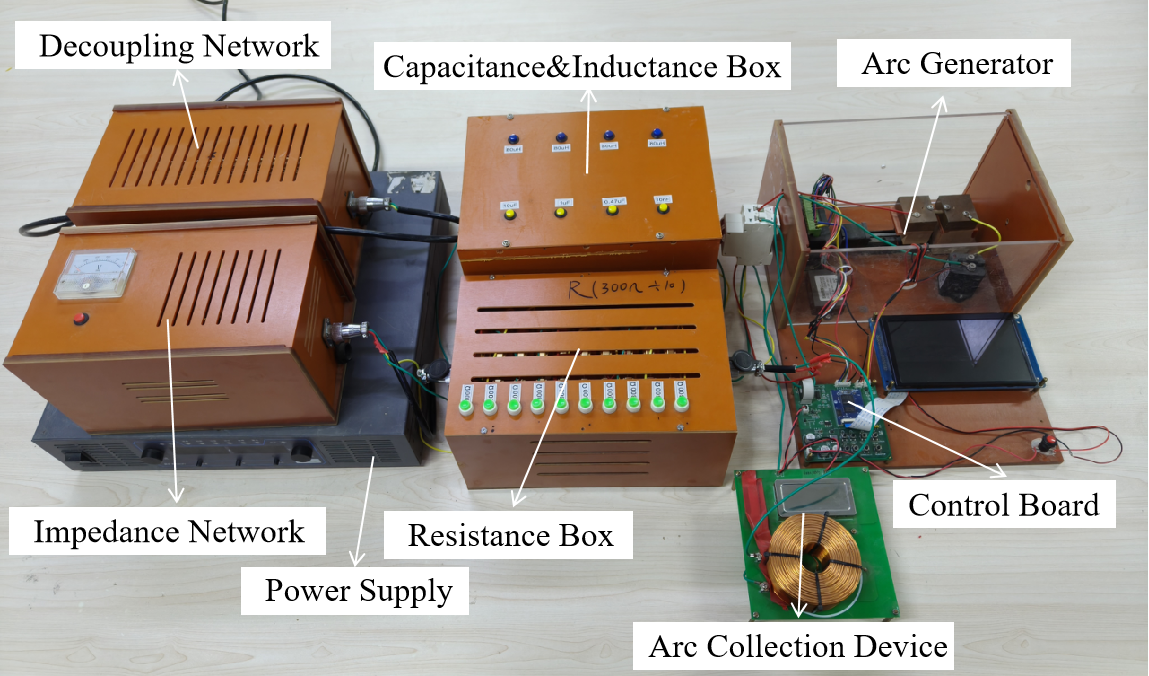}
    \caption{
        Photovoltaic DC Series Arc Fault Experimental Platform.
    }
    \label{fig:Experimental_Platform}
\end{figure}

Arc faults generate high temperatures and electrical sparks. If the faulty circuit is not quickly disconnected, it may lead to fires, and prolonged arcs can damage photovoltaic devices such as inverters, shortening their lifespan and increasing maintenance and replacement costs. Many studies have been conducted to address the problem of arc fault detection~\cite{uriarte2012dc,gammon2014review,khamkar2020arc,yan2023simplified}.

Traditional arc fault detection methods mainly include machine learning methods, time-domain-based methods~\cite{hastings2012direct}, frequency-domain-based methods~\cite{gu2019design,chae2016series}, and time-frequency-domain-based methods~\cite{liu2019application,wang2015arc,chen2019wavelet}. Machine learning methods include Markov models~\cite{telford2016diagnosis}, ant colony algorithms~\cite{chen2022feature}, entropy models~\cite{georgijevic2015detection}, and Support Vector Machines (SVM)~\cite{li2022research}. Time-domain-based methods are limited in their inability to capture frequency characteristics, while frequency-domain-based methods require Fourier transformation of the current signal to compare the amplitude-frequency characteristics, but they lack time localization ability. Time-frequency-domain-based methods can comprehensively analyze time-frequency characteristics, but they suffer from limitations in generalizability and robustness.

\begin{figure}[t]
    \centering
    \includegraphics[width=8.5cm]{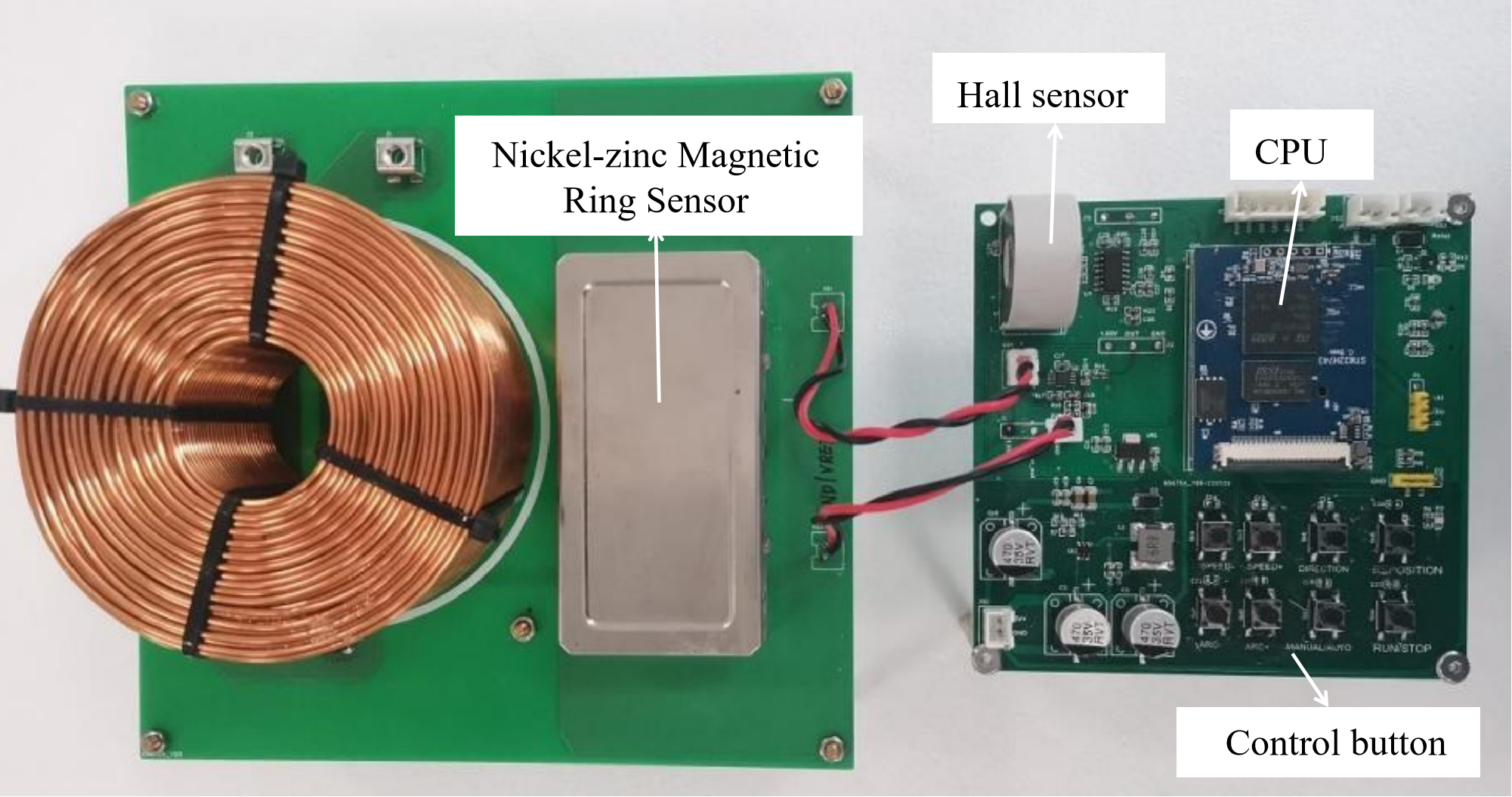}
    \caption{
        Sensor and motherboard design of the Arc Collection Device.
    }
    \label{fig:sensor}
\end{figure}

Deep learning-based methods mainly include Convolutional Neural Networks (CNN)~\cite{yang2019novel}, Recurrent Neural Networks (RNN)~\cite{zaremba2014recurrent}, and RNN variants such as Long Short-Term Memory (LSTM)~\cite{hochreiter1997long}. CNNs have limited ability to capture temporal features, and as the network depth increases, they require higher computational resources. In comparison, RNNs and their variants require longer training times and are difficult to parallelize.

Transformer-based methods~\cite{chabert2023transformer, tian2024sunspark} can effectively handle long time-series data and capture arc fault signals more comprehensively. However, Transformer models rely on the Attention mechanism, which, while effective, is inefficient. Therefore, using Transformer models requires balancing the improvements in results with the increased risk of higher response times.

\begin{figure*}[t]
    \centering
    \includegraphics[width=18cm]{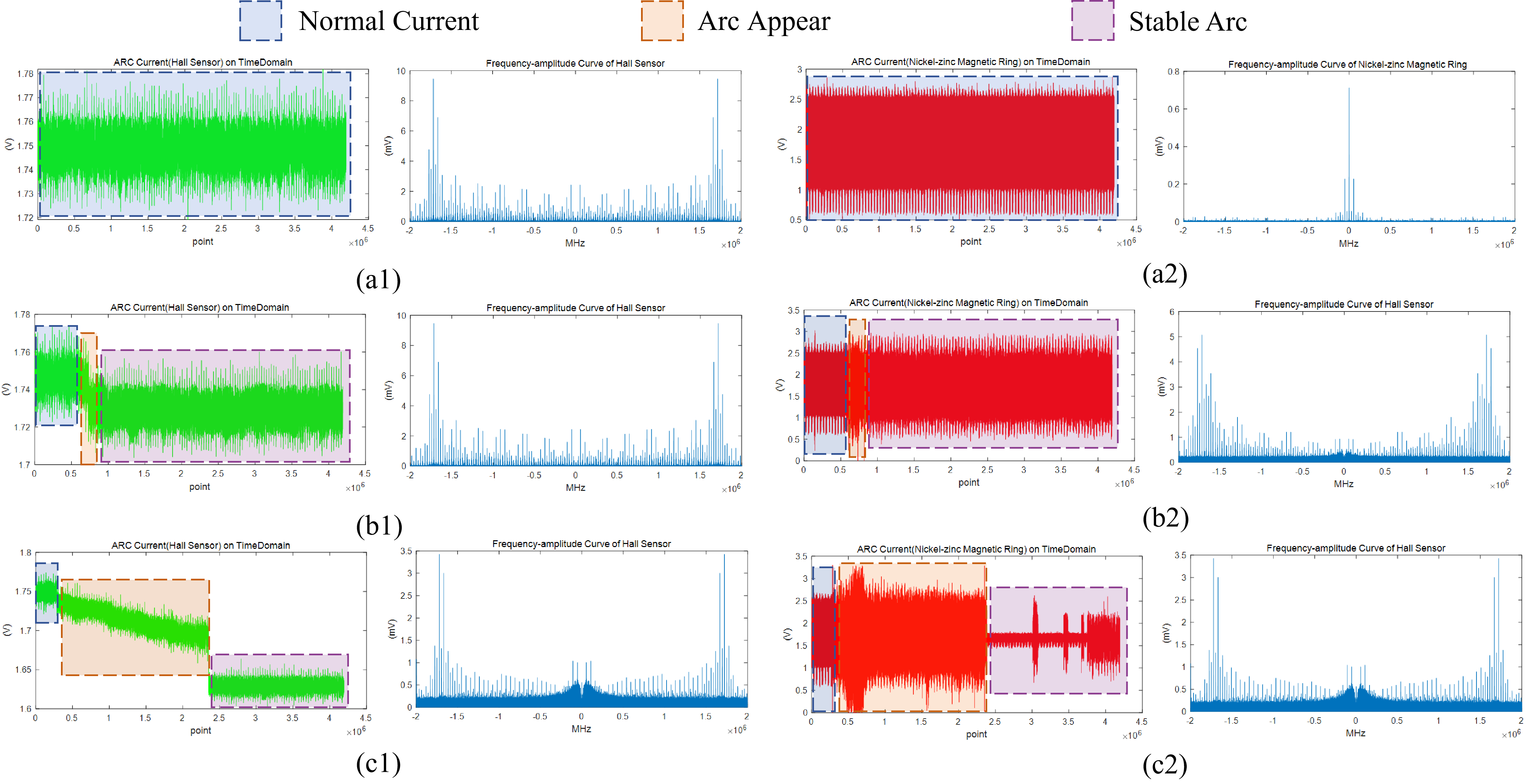}
    \caption{
The figure shows arc fault data obtained under different arc gap and arc speed conditions with a voltage of 100V. The first row (a1)(a2) represents the observations at 100V, with an arc gap of 0.1$mm$ and an arc speed of 0.1$mm/s$, where no arc occurs. The second row (b1)(b2) shows the results at 100V, with an arc gap of 0.2$mm$ and an arc speed of 0.8$mm/s$, where an arc fault occurs. The third row (c1)(c2) represents the observations at 100V, with an arc gap of 1.6$mm$ and an arc speed of 1.6$mm/s$, where an arc fault occurs.
    }
    \label{fig:orange_data} 
\end{figure*}

In this paper, we propose DCAMamba, the first method to apply the Mamba model~\cite{gu2023mamba} to Arc Fault Detection. Specifically, DCAMamba is a novel architecture for arc fault detection. Our framework is based on the State Space Model (SSM)~\cite{gu2021efficiently}. To meet the fast response requirements of arc fault detection~\ref{fig:time}, we employ a hardware-aware algorithm to accelerate the computation process. We replace the convolution in CNNs or the attention mechanism in Transformers with a selective scanning strategy to capture global features, which is both efficient and effective. Additionally, to adapt to arc fault detection, we have modified the partitioning strategy and incorporated a DCAMamba classification head.

After implementing the DCAMamba model, we also propose a Feature Amplification Strategy (FAS) tailored for arc fault signals. As is well known, arc fault distributions are generally more discrete, while normal current distributions tend to be more compact. However, the fluctuations in arc fault data diminish this distribution difference, making it visually challenging to distinguish and posing a difficulty for accurate identification by the model. One intuitive approach is to use Attention to capture global features~\cite{fu2019dual}, but this significantly increases the inference time of the model. To address this, we introduce FAS, which processes the data based on the magnitude of each data point to create a distribution that is easier for the model to recognize. This elegant and efficient method achieves accurate identification while maintaining the fast inference speed of the original model.

In summary, we first apply FAS to process the current signals, and then input the processed results into our designed DCAMamba, ultimately achieving accurate and fast arc fault detection. Our contributions can be summarized as follows:

\begin{itemize}
\item In accordance with the UL1699B standard, we build a photovoltaic DC arc fault experimental platform. Within the voltage range of 100V to 300V, arc and non-arc samples were collected for different types of loads, creating a dataset of arc current under multiple loads. 

\item We propose DCAMamba, which is the first framework based on the Mamba model for arc fault detection. We optimized the chunking strategy and modified the model structure to better suit the arc fault detection task.

\item In response to the general distribution of arc fault data, we propose the Feature Amplification Strategy (FAS) to process the data. This strategy significantly enhances performance while maintaining faster response times.

\item The experimental results demonstrate that we can achieve fast response and accurate arc fault detection. The method achieved an accuracy of 96.72\% on the test set, with an inference speed of only 1.87 milliseconds. The extremely fast response time makes DCAMamba suitable for industrial applications and positions it as a potential foundational model for future arc fault detection methods.

\end{itemize}

\section{ARC DATA ACQUISITION AND ANALYSIS }
\label{sec:ArcData}

\subsection{Construction of the Experimental Platform}
Due to the wide frequency spectrum and weak signal characteristics of DC arc fault currents, more refined data acquisition techniques and signal processing methods are required, which presents a challenge for arc fault detection. In this section, we built a photovoltaic DC series arc fault experimental platform based on the UL1699B standard for collecting arc fault data. Compared with traditional arc fault collection methods, this platform offers faster data collection speeds, as shown in Figure~\ref{fig:Experimental_Platform}. The experimental platform includes a DC power supply, a DC arc signal collector, a DC arc generator, a decoupling network, an impedance network, and a load. The arc signal collector, shown in Figure~\ref{fig:sensor}, consists of an STM32H743XI microcontroller and a high-sensitivity arc signal collection front end based on the law of electromagnetic induction. It can automatically adjust the arc gap and arc speed. We set the sampling rate of the arc collector to 4MSPS, with a sampling resolution of 12 bits for the analog-to-digital conversion (ADC) circuit. After sampling, the raw arc data is visualized, as shown in the figure. Each sampled data is divided into three stages: (a) no arc, (b) gradually igniting arc, and (c) stable burning arc. By adjusting different loads (resistor, capacitor, inductor), voltage ranges (100V, 150V, 200V, 300V), arc speeds (0.7$mm/s$ to 2.6$mm/s$), and arc gaps (0.7$mm$ to 2.6$mm$), we collected fault arc data from various types and scenarios.

\begin{figure*}[t]
    \centering
    \includegraphics[width=18cm]{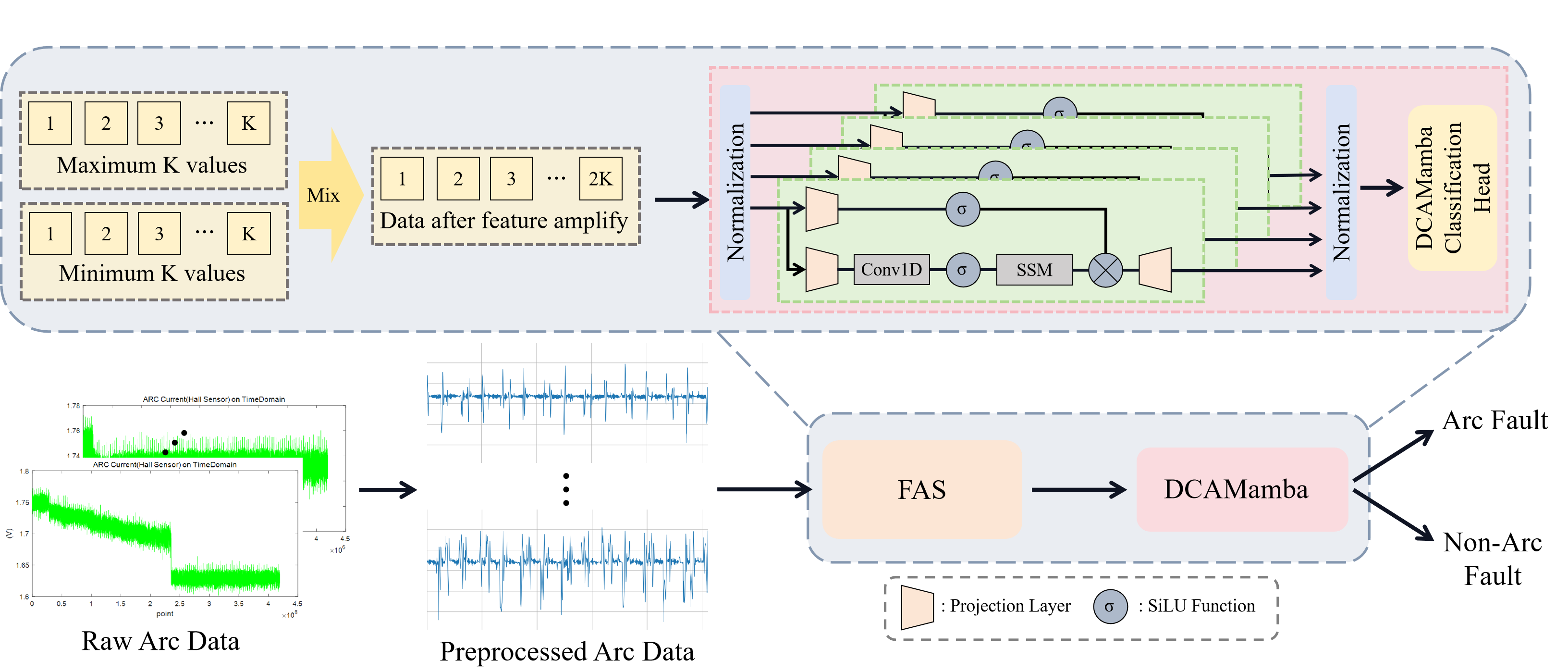}
    \caption{DCAMamba Pipeline: The data obtained from the experimental platform undergoes initial preprocessing to distinguish between fault and normal data. The processed data is then restructured by the FAS module. Next, the data is input into the DCAMamba model, where it is normalized before entering the model blocks. After computation in the Mamba Block, another normalization layer is applied, and the data is then passed to the classification head to produce the final result.}
    \label{fig:main} 
\end{figure*}

\subsection{Arc Fault Dataset Construction}
First, we visualized the collected raw data, as shown in the Figure~\ref{fig:orange_data}. We observed that each raw data sequence consists of three stages: the stage without arc occurrence, the stage where the arc begins to form, and the stage where the arc stabilizes. Present the time-domain and frequency-domain characteristics of the arc signals collected under three different conditions using both Ni-Zn ferrite core sensors and Hall sensors. In the first set of experiments (100V, arc gap 0.1$mm$, arc speed 0.1$mm/s$), no arc occurred. The time-domain signal was stable, and the frequency-domain signal exhibited concentrated energy with simple features. In the second set (100V, arc gap 0.2$mm$, arc speed 0.8$mm/s$) and third set (100V, arc gap 1.6$mm$, arc speed 1.6$mm/s$), arcs occurred. At this point, the time-domain signal amplitude increased significantly, particularly in the third set, where the signal exhibited violent fluctuations. The energy distribution of the frequency-domain signal became more complex, with a notable increase in high-frequency components.

\begin{figure*}[t]
    \centering
    \includegraphics[width=15cm]{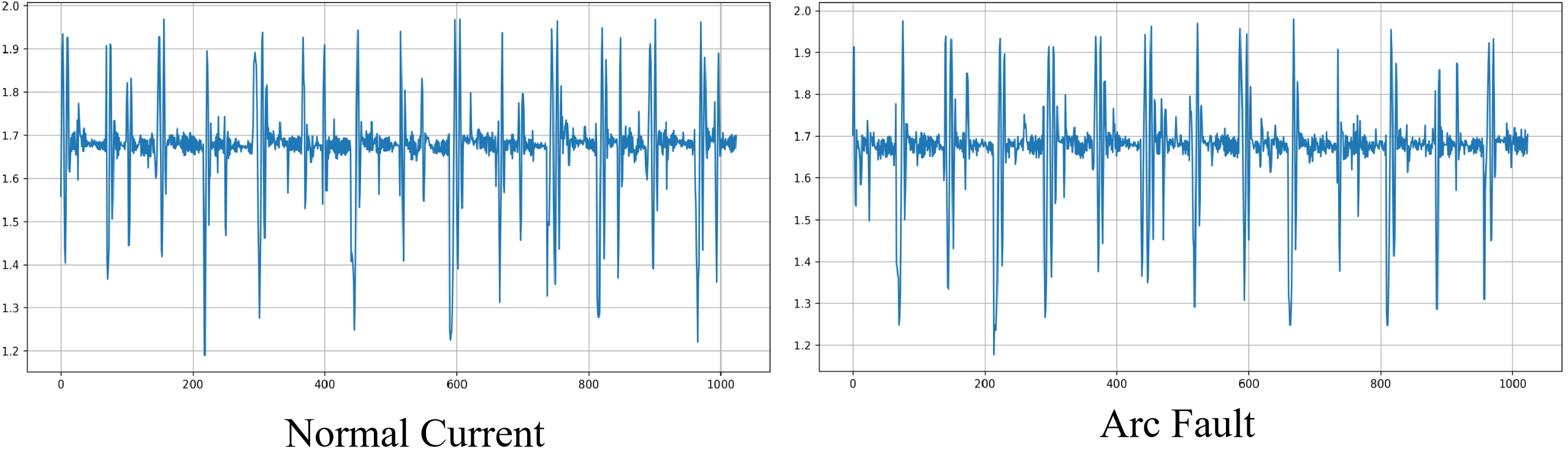}
    \caption{
   Visualization of the preprocessed Normal Current and Arc Fault data.
    }
    \label{fig:Current} 
\end{figure*}

By comparing the response characteristics of the two types of sensors, the Ni-Zn ferrite core sensor (red) is more sensitive in capturing the transient characteristics and high-frequency signals of arc discharge, while the Hall sensor (green) is more suitable for describing the overall trend of signal variations. To better facilitate model training, we need to segment the raw large data into subsequences representing both non-arc and arc occurrences. Each subsequence has a length of 1024 data points. The segmented subsequences not only retain the details and accuracy of the original data but also effectively reduce the complexity of subsequent arc data during neural network training. Finally, we visualized the 1024 data points of each subsequence, as shown in Figure~\ref{fig:Current}. Then, we analyzed the arc subsequences from a distribution perspective and used mathematical methods to model the data. We found that normal current tends to have a more concentrated distribution, while arc fault data tends to have a more scattered distribution. This finding is crucial for our subsequent research and modeling efforts.

\section{DCAMamba Training Process}
\label{sec:DCAMamba Training Process}
Arc fault detection is essentially a classification task. By using the SSM model, a dynamic system modeling approach based on state-space representation, DCAMamba can effectively capture the current features in arc fault detection. In Section 3.1, we first introduce the SSM model; in Section 3.2, we present the theoretical basis and implementation of the Feature Amplification Strategy (FAS), in Section 3.3, we provide an overview of the design and implementation of DCAMamba.

\begin{figure}[t]
    \centering
    \includegraphics[width=8.5cm]{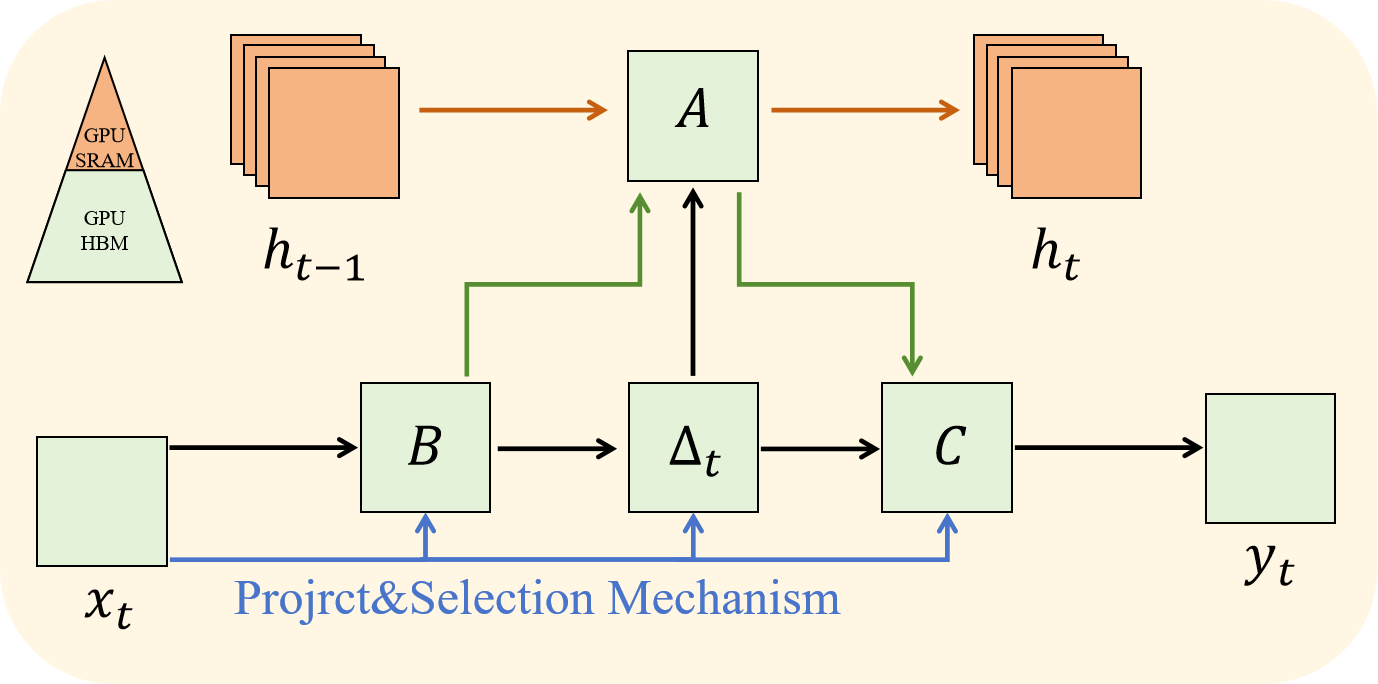}
    \caption{The structured SSM architecture maps each input channel to the output, capturing the dynamic characteristics of the current signals. The selection mechanism introduces input-dependent dynamics, while the hardware-aware algorithm enables more efficient implementation of the extended state in the GPU memory hierarchy.}
    \label{fig:ssm} 
\end{figure}

\subsection{Preliminary}
\label{sec:Preliminary}
The State Space Model (SSM) is a mathematical framework used to describe dynamic systems, originating from the Kalman filter~\cite{kalman1960new}, and can be considered a linear time-invariant system. A classical state space model constructs two key equations: the state equation and the observation equation. It models the relationship between the input $x(t) \in \mathbb{R}^N$ and the output $y(t) \in \mathbb{R}^N$ at the current time $t$ using an $N$-dimensional hidden state $h(t) \in \mathbb{R}^N$. This process can be described by a linear ordinary differential equation (ODE).

\begin{equation}
\begin{aligned}
    &h^{\prime}(t) = \mathbf{A}h(t) + \mathbf{B}x(t)\\
    &y(t) = \mathbf{C}h(t)
\end{aligned}
\end{equation}

Where $\mathbf{A} \in \mathbb{R}^{N \times N}$ is the state transition matrix, $\mathbf{B} \in \mathbb{R}^N$ and $\mathbf{C} \in \mathbb{R}^N$ are the projection matrices. The three continuous parameters $y \in \mathbb{R}^N$ are computed from the input $x \in \mathbb{R}^N$ and the hidden state $h \in \mathbb{R}^N$. To integrate the continuous-time State Space Model (SSM) into a deep learning model, discretization is required. Here, $\mathbf{A}$ and $\mathbf{B}$ are discretized using a zero-order hold (ZOH) method with a time scale parameter $\Delta$. The process is as follows.

\begin{equation}
\begin{aligned}
\overline{\mathbf{A}} &= \exp(\Delta \mathbf{A}) \\
\overline{\mathbf{B}} &= (\Delta \mathbf{A})^{-1} \left( \exp(\Delta \mathbf{A}) - \mathbf{I} \right) \cdot \Delta \mathbf{B}
\end{aligned}
\end{equation}
After discretizing the continuous parameters, equation (1) can be rewritten as:
\begin{equation}
\begin{aligned}
h_k &= \overline{\mathbf{A}} h_{k-1} + \overline{\mathbf{B}} x_k \\
y_k &= \mathbf{C} h_k
\end{aligned}
\end{equation}
Finally, the input of the computation can be represented using a convolution, as shown below:
\begin{equation}
\begin{aligned}
\overline{\mathbf{K}} &= \left( \mathbf{C} \overline{\mathbf{B}}, \, \mathbf{C} \overline{\mathbf{A}} \overline{\mathbf{B}}, \, \dots, \, \mathbf{C} \overline{\mathbf{A}}^{L-1} \overline{\mathbf{B}} \right) \\
y &= x * \overline{\mathbf{K}}
\end{aligned}
\end{equation}
Where \( L \) is the length of the input sequence, and \( \overline{\mathbf{K}} \in \mathbb{R}^L \) represents the structured convolution kernel.

\subsection{Feature Amplification Strategy}
\label{sec:Feature Amplification Strategy}
In this section, we will introduce the Feature Amplification Strategy (FAS) in detail, which is a data processing strategy specifically for arc fault detection. However, it is fundamentally different from simple data preprocessing methods. Specifically, previous arc fault detection methods often treat arc data as non-stationary, nonlinear transient signals with specific pattern characteristics. Although these methods can achieve arc fault detection, we found that distinguishing between arc and non-arc signals is challenging due to the oscillatory distribution of the arc data. As shown in Figure~\ref{fig:Current}, in the middle range of the current signal (i.e., when the current is 1.7A), the signal amplitude exhibits periodic repetition within a relatively stable interval. This increases the challenge of distinguishing between the two classes and has an adverse impact on the learning process of the model. We re-analyzed this problem from a novel perspective by modeling the data distribution of normal current and abnormal current. We used Matlab to analyze the distribution patterns of the two sets of data and concluded that normal current tends to exhibit a compact distribution, while abnormal current tends to exhibit a more dispersed distribution. To address this phenomenon, we proposed the FAS method, which is a simple yet effective strategy.

We represent the preprocessed data as \( x \in \mathbb{R}^{B \times C \times L} \), where \( B \) is the batch size, \( C \) is the number of channels, and \( L \) is the length of each sequence. To extract important values from each sequence, we perform the following operation. For each batch \( b \in \{1, \dots, B\} \) and each channel \( c \in \{1, \dots, C\} \), we calculate the top \( K \) values as follows:

\begin{equation}
    \mathbf{V}_{\text{Max}}^{(b,c)} = \text{TopK}_{\text{L}} (x^{(b,c)}, K)
\end{equation}
Where \( \mathbf{V}_{\text{Max}}^{(b,c)} \in \mathbb{R}^K \) contains the top \( K \) values.
Similarly, the process of calculating the bottom \( K \) values can be described as follows:

\begin{equation}
    \mathbf{V}_{\text{Min}}^{(b,c)} = \text{TopK}_{\text{S}} (x^{(b,c)}, K)
\end{equation}
Where \( \mathbf{V}_{\text{Min}}^{(b,c)} \in \mathbb{R}^K \) represents the bottom \( K \) values.

We concatenate the obtained \( \mathbf{V}_{\text{Max}}^{(b,c)} \) and \( \mathbf{V}_{\text{Min}}^{(b,c)} \) to form a composite tensor:

\begin{equation}
    \mathbf{V}_{\text{Com}}^{(b,c)} = 
    \begin{bmatrix}
        \mathbf{V}_{\text{Max}}^{(b,c)} \\ 
        \mathbf{V}_{\text{Min}}^{(b,c)}
    \end{bmatrix}
    \in \mathbb{R}^{2K}
\end{equation}

The resulting tensor \( \mathbf{V}_{\text{Com}} \in \mathbb{R}^{B \times C \times 2K} \) aggregates the most significant distribution results from the input sequence. With the FAS strategy, we not only extract features from the data more effectively but also further prune the data, thereby improving the inference speed of the model.

\subsection{DCAMamba}
\label{sec:DCAMamba}
In Figure~\ref{fig:main}, we present the full pipeline of DCAMamba. Inspired by the recent Mamba model~\cite{gu2023mamba} proposed for text-related tasks, we have adopted its design, inheriting the hardware-aware algorithm and scanning mechanism from Mamba. Additionally, we adjusted the chunking strategy and data processing methods of the model specifically for arc fault detection, and ultimately incorporated the DCAMamba Classification Head to enable arc fault detection.

Specifically, in DCAMamba, we removed the text embedding method originally used in Mamba to better suit the data structure of arc fault detection. Although directly feeding arc data into the model might seem like a straightforward approach, the irregular waveform distribution of current data would lead the model to learn a lot of irrelevant features, which negatively affects detection accuracy. Therefore, we introduced the Feature Amplification Strategy (FAS), which replaces the text embedding method. The specific implementation of FAS is described in \ref{sec:Feature Amplification Strategy}. We denote the original current input as $X \in \mathbb{R}^{B,C,L}$, and after processing with FAS, it is normalized to $x \in \mathbb{R}^{B,C,2K}$, which is then passed through multiple DCAMamba blocks for feature extraction. Each DCAMamba block includes operations such as input projection, 1D convolution, activation functions, and output projection to extract deeper features from the arc fault data.

Next, we detail the inference process of the DCAMamba block. First, the normalized result $x \in \mathbb{R}^{B,C,2K}$ is projected into the $2ED$ space through a linear layer:

\begin{equation}
    x' = x \cdot \mathbf{W}_{\text{in}} + \mathbf{b}_{\text{in}}
\end{equation}
where $\quad x' \in \mathbb{R}^{B \times C \times 2ED}$. The projected tensor is split into two parts, $x_1$ and $z$, and $x_1$ undergoes deep convolution, activation, and SSM operations:

\begin{equation}
    y = \text{SSM} \left( \sigma \left( \text{Conv1D} \left( x_1^T \right)^T\right) \right)
\end{equation}
where $\quad y \in \mathbb{R}^{B \times C \times ED}$, and $x_1^T$ denotes the transpose of $x_1$ from shape $(B, C, ED)$ to $(B, ED, C)$ to fit the convolution operation; $\text{Conv1D}$ refers to the deep convolution operation that maintains the sequence length; $\sigma$ represents the SiLU activation function; $\text{SSM}$ refers to the Selective Scanning Module. Apply the SiLU activation function to $z$:

\begin{equation}
    z' = \sigma(z)
\end{equation}
where $\quad z' \in \mathbb{R}^{B \times C \times ED}$. Then, merge the two branches by element-wise multiplying the output $y$ from the convolutional branch and $z'$:

\begin{equation}
    \mathbf{o} = y \odot z'
\end{equation}
where $\quad \mathbf{o} \in \mathbb{R}^{B \times C \times 2ED}$. Project the merged output back to the original model dimension $D$ using a linear layer:

\begin{equation}
    y_{\text{out}} = \mathbf{o} \cdot \mathbf{W}_{\text{out}} + \mathbf{b}_{\text{out}}
\end{equation}
where $\quad y_{\text{out}} \in \mathbb{R}^{B \times C \times 2K}$. Finally, after the computation through the block, the model maps the features to a binary classification output using a linear classification layer. The entire architecture is normalized using RMSNorm to ensure stable training.

\begin{table*}[t]
\centering
\caption{Comparison of experimental results with mainstream models on the following metrics: Precision$(\%)\uparrow$, Recall$(\%)\uparrow$, F1$(\%)\uparrow$, training loss Train\_loss$(\%)\downarrow$, validation accuracy Val\_acc$(\%)\uparrow$, validation loss Val\_loss$(\%)\downarrow$, and inference time IT$(ms)\downarrow$.}
\label{table:Comparison Experiment}
\begin{tabular}{l c c c c c c c}
\hline
\textbf{Models}     & \textbf{Precision$(\%)\uparrow$} & \textbf{Recall$(\%)\uparrow$} & \textbf{F1$(\%)\uparrow$}    & \textbf{Train\_loss$(\%)\downarrow$} & \textbf{Val\_acc$(\%)\uparrow$} & \textbf{Val\_loss$(\%)\downarrow$} & \textbf{IT $(ms)\downarrow$} \\ \hline 
CNN                 & 87.81             & 85.41           & 85.18          & 0.3824               & 85.42             & 0.3328             & 2.52\\ 
RNN                 & 96.22             & 96.12           & 96.12          & 0.081                & 96.12             & \textbf{0.093}              & 11.88 \\ 
RCNN                & 89.69             & 89.68           & 89.68          & 0.016                & 89.68             & 0.3148             & 17.48\\ 
RNN-ATT             & 89.90             & 89.74           & 89.44          & 0.229                & 89.74             & 0.189              & 12.19\\  
Transformer         & 90.74             & 90.72           & 90.72          & 0.1361               & 90.72             & 0.2056             & 16.54 \\ 
Mamba               & 86.27             & 86.28           & 86.27          & \textbf{0.013}                & 86.23             & 0.3511             & 1.90\\ 
DCAMamba (Ours)     & \textbf{96.72}    & \textbf{96.71} & \textbf{96.72}  & 0.015    & \textbf{96.72}   & 0.1112        & \textbf{1.87} \\ \hline
\end{tabular}
\end{table*}

\section{EXPERIMENTAL RESULTS AND ANALYSES}
\label{sec:experiment}

\subsection{Implementation Setup}

\subsubsection{Arc Fault Dataset}
We collected arc fault data using a self-sampling device based on the principle of electromagnetic induction. Since the generation of fault arcs is closely related to factors such as load type, voltage range, arc pull speed, and arc gap, we designed various experimental scenarios for data collection. To better replicate real-world application environments, we selected resistive, capacitive, and inductive loads as a combination for multi-scenario simulations. This approach allows for a more accurate reproduction of real-world conditions.

Arc fault data was collected within different voltage ranges (100V to 300V), covering various scenarios from low to high voltage. Additionally, different arc pull speeds (0.7 to 2.6) and arc gaps (0.7 to 2.6) were tested. The results showed that at lower pull speeds, the arc was stable and lasted longer, while at higher speeds, the arc became unstable with more severe waveform fluctuations. Larger arc gaps required higher voltages to sustain the arc, increasing the difficulty of fault detection.

The device captured current data at a high frequency of 4 million samples per second under different voltage and load conditions. Each arc pull lasted one second, meaning each record contained 4 million data points. Due to the large volume of data in each record, the computational complexity for the model increased, which clearly does not meet the requirements for accurate and fast arc fault detection. To address this issue, we sliced each initial dataset into segments of 1024 data points. This approach preserves the accuracy of the original data while reducing the computational burden on the model.

\begin{figure}[t]
    \centering
    \includegraphics[width=8.5cm]{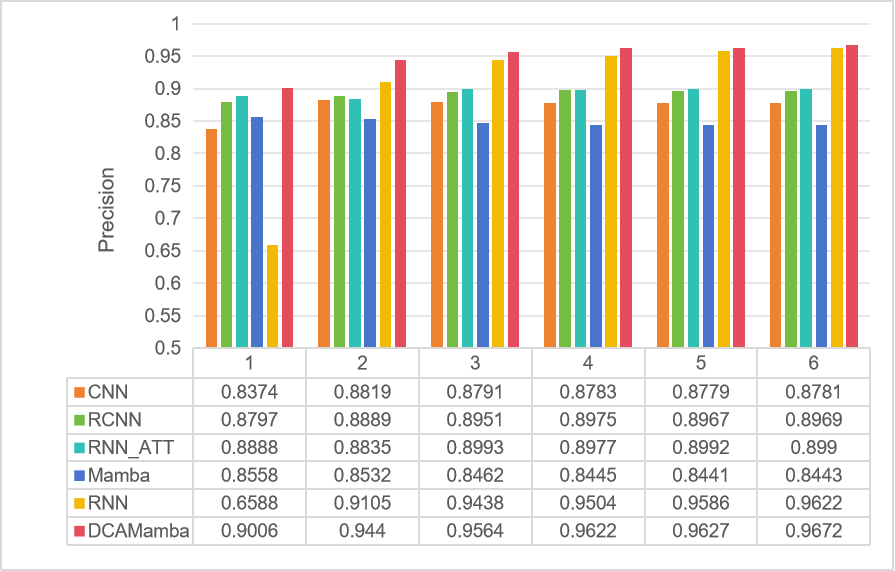}
    \caption{Comparison of experimental results with other mainstream methods in terms of Precision for different epochs.}
    \label{fig:precision} 
\end{figure}

\begin{figure}[t]
    \centering
    \includegraphics[width=8.5cm]{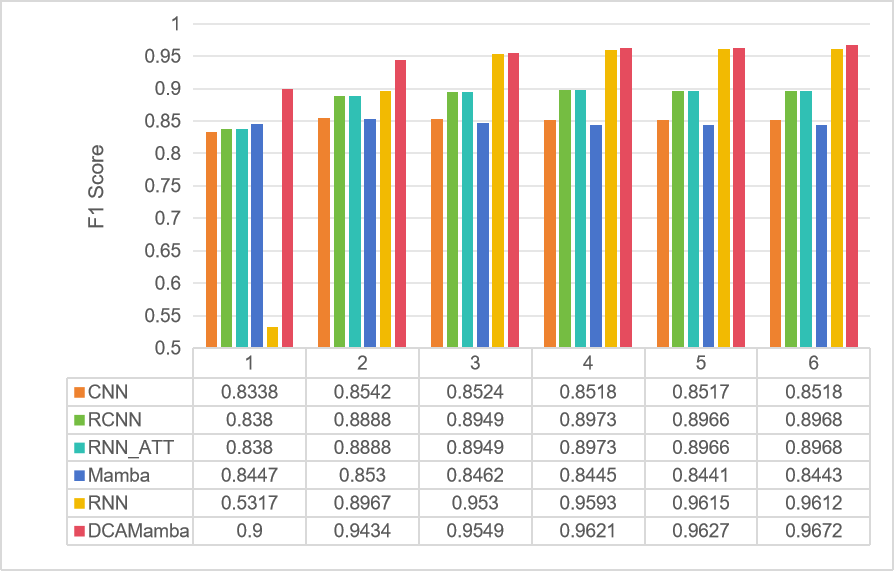}
    \caption{Comparison of experimental results with other mainstream methods in terms of F1 Score for different epochs.}
    \label{fig:f1_score} 
\end{figure}

\begin{figure}[t]
    \centering
    \includegraphics[width=8.5cm]{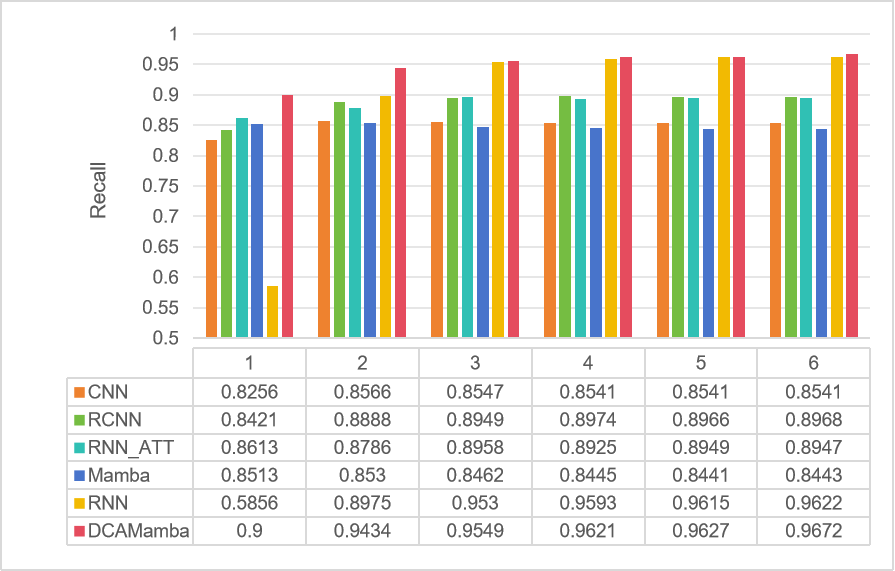}
    \caption{Comparison of experimental results with other mainstream methods in terms of Recall for different epochs.}
    \label{fig:recall} 
\end{figure}

\subsubsection{Experimental Setup for DCAMamba}
After obtaining the sliced datasets, we applied the Feature Amplification Strategy (FAS) to further process the arc fault data, extracting feature data that represents the distribution of different categories of arc fault data. The FAS-processed data was then input into the DCAMamba model for training. During the training process, the arc fault data first passes through a normalization layer before entering multiple Blocks within the model. Each Block processes the features through operations such as normalization, convolution, activation, and projection. Finally, the classification head of the DCAMamba model outputs a binary classification result to determine whether an arc fault has occurred.

Our experiments were conducted on a single NVIDIA RTX 4090 GPU. The collected data in the dataset was divided into a training set and a test set in a 7:3 ratio. During training, an adaptive learning rate strategy was used for 100 epochs, with a batch size of 128 and a learning rate of $1e-4$. The optimizer used was Adam, the activation function was ReLU, and the loss function was cross-entropy loss.

\subsubsection{Evaluation Metrics}
In deep learning-based arc fault detection tasks, commonly used evaluation metrics include Accuracy, Recall, Precision, and F1 Score, which assess model performance from different perspectives. Accuracy measures the proportion of correctly predicted samples out of the total number of samples and is used for overall performance evaluation. Recall focuses on the ability of the model to capture positive samples, specifically the proportion of correctly identified positive samples out of all actual positive samples, reflecting sensitivity to false negatives. Precision measures the proportion of true positive samples among those predicted as positive by the model, reflecting the ability of the model to control false positives. The F1 Score is the harmonic mean of Precision and Recall, used to find a balance between them, particularly suited for evaluating model performance in imbalanced class situations.

\subsection{Comparison with Other Models}
To comprehensively evaluate the performance of the proposed DCAMamba model, we conducted a series of comparative experiments, systematically comparing it with several classic and state-of-the-art models. The comparison models include the traditional Convolutional Neural Network (CNN)~\cite{kim2014convolutional}, Recurrent Neural Network (RNN)~\cite{zaremba2014recurrent}, the combination of CNN and RNN (RCNN)~\cite{girshick2014rich}, RNN with Attention Mechanism (RNN-ATT)~\cite{jin2017multimodal}, Transformer~\cite{vaswani2017attention}, and the previously mentioned Mamba~\cite{gu2023mamba} model.

The comparison experiments cover several key metrics to assess the performance of different models comprehensively. Table~\ref{table:Comparison Experiment} shows the performance of various models on multiple evaluation metrics, including Precision, Recall, F1 Score, Training Loss, Validation Accuracy, Validation Loss, and Inference Time (IT). The experimental results demonstrate that the proposed DCAMamba model outperforms all other models in every metric.

Specifically, the DCAMamba model achieves a Precision, Recall, and F1 Score of 96.72\%, surpassing all comparison models. It also has a training loss of 0.015, validation loss of 0.1112, and a validation accuracy of 96.72\%, demonstrating strong generalization capability. Additionally, the DCAMamba model has an inference time of 1.87 $ms$, significantly outperforming most other models (e.g., RCNN and Transformer), showcasing higher computational efficiency.

In contrast, traditional deep learning models (such as CNN and RCNN) perform relatively poorly. The CNN achieves only 87.81\% in accuracy, with an F1 Score of 85.18\%, and an inference time of 2.52 $ms$. RNN and its variant (RNN-ATT) show some improvement in performance, but they are too slow for industrial applications. The Transformer model performs similarly to DCAMamba, but its inference time is significantly longer (16.54 $ms$). Compared to the Mamba model, DCAMamba improves accuracy by 10.45\% while maintaining the fast response inherited from Mamba.

To provide a more comprehensive and intuitive demonstration of the performance of these methods, we compared the aforementioned approaches across multiple training epochs using three metrics: precision (Figure\ref{fig:precision}), F1 score (Figure\ref{fig:f1_score}), and recall (Figure\ref{fig:recall}). In each of the three graphs, the labels 1 through 6 correspond to the results obtained at epochs 1, 20, 40, 60, 80, and 100, respectively. The results clearly show that both our method and the RNN method exhibit significant advantages, with our approach consistently outperforming RNN across all metrics and epochs. Additionally, as illustrated in Figure\ref{fig:time}, our method achieves a 6.5-fold improvement in performance compared to the RNN approach.

In conclusion, the DCAMamba model not only ensures high accuracy, high recall, and low loss but also significantly improves inference efficiency, demonstrating its potential for practical arc fault detection applications.

\begin{table}[t]
\centering
\caption{Ablation experiments of feature extraction parameters on Precision$(\%)\uparrow$and response time IT $(ms)\downarrow$indexes.}
\label{table:feature_extraction}
\begin{tabular}{l c c }
\hline
\textbf{Feature extraction parameter}     & \textbf{Precision$(\%)\uparrow$}   & \textbf{IT $(ms)\downarrow$} \\ \hline 
Mamba               & 84.51                           & 1.90\\ 
$K$=128                 & 94.45                           & \textbf{1.80}\\ 
$K$=256                 & 96.02                           & 1.83\\ 
$K$=512                 & \textbf{96.72}                           & 1.87 \\ \hline 
\end{tabular}
\end{table}

\begin{table}[t]
\centering
\caption{AAblation experiments on the number of blocks in DCAMamba, examining Precision$(\%)\uparrow$and response time IT $(ms)\downarrow$indexes.}
\label{table:block_experiment}
\begin{tabular}{l c c }
\hline
\textbf{Number of Blocks}     & \textbf{Precision$(\%)\uparrow$}   & \textbf{IT $(ms)\downarrow$} \\ \hline 
2                     & 96.38                           & \textbf{1.13}\\ 
4                     & \textbf{96.72}                           & 1.87\\ 
8                     & 94.84                           & 3.36\\ 
16                    & 95.09                           & 5.96 \\ \hline 
\end{tabular}
\end{table}

\begin{table}[t]
\centering
\caption{Experiment on different classification heads of DCAMamba, evaluating 
 Precision$(\%)\uparrow$, Recall$(\%)\uparrow$and response time IT $(ms)\downarrow$.}
\label{table:Classification—Head}
\begin{tabular}{l c c c}
\hline
\textbf{Classification Head}     & \textbf{Precision$(\%)\uparrow$}  & \textbf{Recall$(\%)\uparrow$}  & \textbf{IT $(ms)\downarrow$} \\ \hline 
Ours                   & \textbf{96.72}            & 96.71                    & \textbf{1.87}\\ 
+Dropout               & 96.43            & 96.43                    & 1.88\\ 
+MLP                   & 96.11            & 96.14                    & 1.99\\ 
+Pooling               & 96.53            & 96.52                    & 1.90 \\ \hline 
\end{tabular}
\end{table}

\subsection{Ablation experiment}
\subsubsection{Feature Amplification Strategy parameters}
To achieve more accurate arc fault detection, we propose the Feature Amplification Strategy. To investigate the impact of different feature extraction parameters $K$ on model performance, we conducted an ablation study. Table~\ref{table:feature_extraction} presents the effects of different feature extraction parameters $K$ on model performance. The experimental results indicate that as $K$ increases, the precision of the model (Precision) improves significantly, while the inference time (IT) increases only slightly. Specifically, when using the default Mamba model (without the Feature Amplification module), the precision is only 84.51\%, and the inference time is 1.90$ms$. After introducing the feature extraction parameter $K$, the performance of the model improves significantly. When $K=128$, the precision reaches 94.45\%, with an inference time of 1.80$ms$; when $K=256$, the precision further improves to 96.02\%, with the inference time slightly increasing to 1.83$ms$; when $K=512$, the model achieves the best performance, with a precision of 96.72\% and an inference time of 1.87$ms$.

\subsubsection{DCAMamba Block}
As the number of Blocks in the DCAMamba model significantly affects accuracy and operational efficiency, we conducted an ablation study to evaluate different Block numbers. Table~\ref{table:block_experiment} shows the impact of various Block numbers on model performance, with evaluation based on precision (Precision) and inference time (Inference Time, IT). The experimental results indicate that changes in the number of blocks significantly impact the performance and inference efficiency of the model. When the number of Blocks is 2, the model achieves a precision of 96.38$\%$ with an inference time of only 1.13$ms$, demonstrating high inference efficiency. As the number of Blocks increases to 4, the precision further improves to 96.72$\%$, and the inference time increases slightly to 1.87$ms$, achieving optimal performance. However, as the number of blocks continues to increase to 8 and 16, the precision of the model decreases to 94.84$\%$ and 95.09$\%$, respectively, while the inference time increases significantly, reaching 3.36$ms$ and 5.96$ms$. This indicates that an excessive number of Blocks leads to model overcomplication, thereby reducing performance and significantly increasing inference time. In conclusion, an appropriate number of Blocks (such as 4) achieves the best balance between accuracy and efficiency, ensuring high precision while maintaining a low inference time.

\subsubsection{Different Classification Heads}
To comprehensively evaluate the performance of the model, an ablation study was also conducted on different classification heads. Table~\ref{table:Classification—Head} presents the experimental results for our model with different classification heads, evaluating the performance in terms of precision (Precision), recall (Recall), and inference time (IT). The experimental results indicate that the proposed classification head achieves the highest precision (96.72$\%$) and recall (96.71$\%$), while the inference time is the shortest, at only 1.87$ms$. When using a dropout-based classification head (+Dropout), both precision and recall slightly decrease to 96.43$\%$, and the inference time slightly increases to 1.88$ms$. The performance is worst when using a multilayer perceptron as the classification head (+MLP), with precision and recall dropping to 96.11$\%$ and 96.14$\%$, respectively, and inference time increasing to 1.99$ms$. Additionally, using a pooling-based classification head (+Pooling) achieves a balance between performance and efficiency, with precision and recall of 96.53$\%$ and 96.52$\%$, respectively, and an inference time of 1.90$ms$. Overall, the experimental results clearly demonstrate that the proposed classification head achieves the best balance between precision and efficiency.

\begin{figure}[t]
    \centering
    \includegraphics[width=8.5cm]{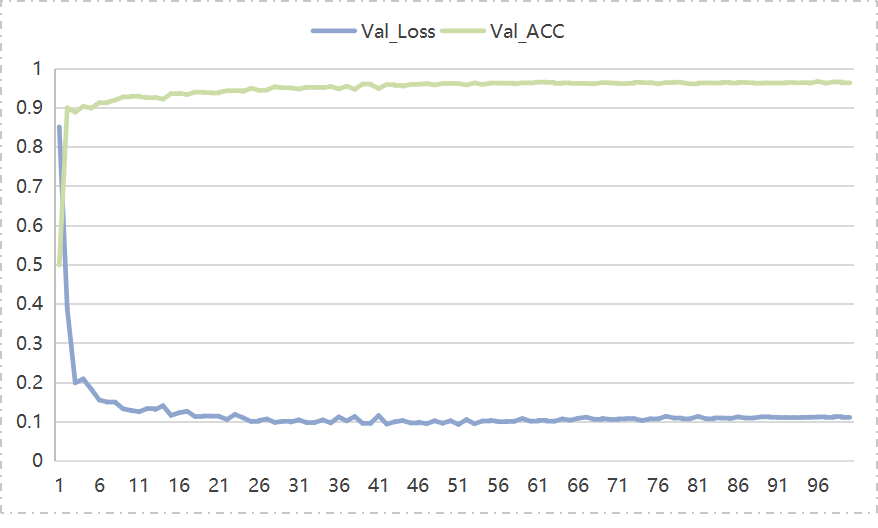}
    \caption{A visual process for verifying loss and verifying changes in the accuracy rate with epochs 1 to 100.}
    \label{fig:val_loss_acc}
\end{figure}

\begin{figure}[t]
    \centering
    \includegraphics[width=8.5cm]{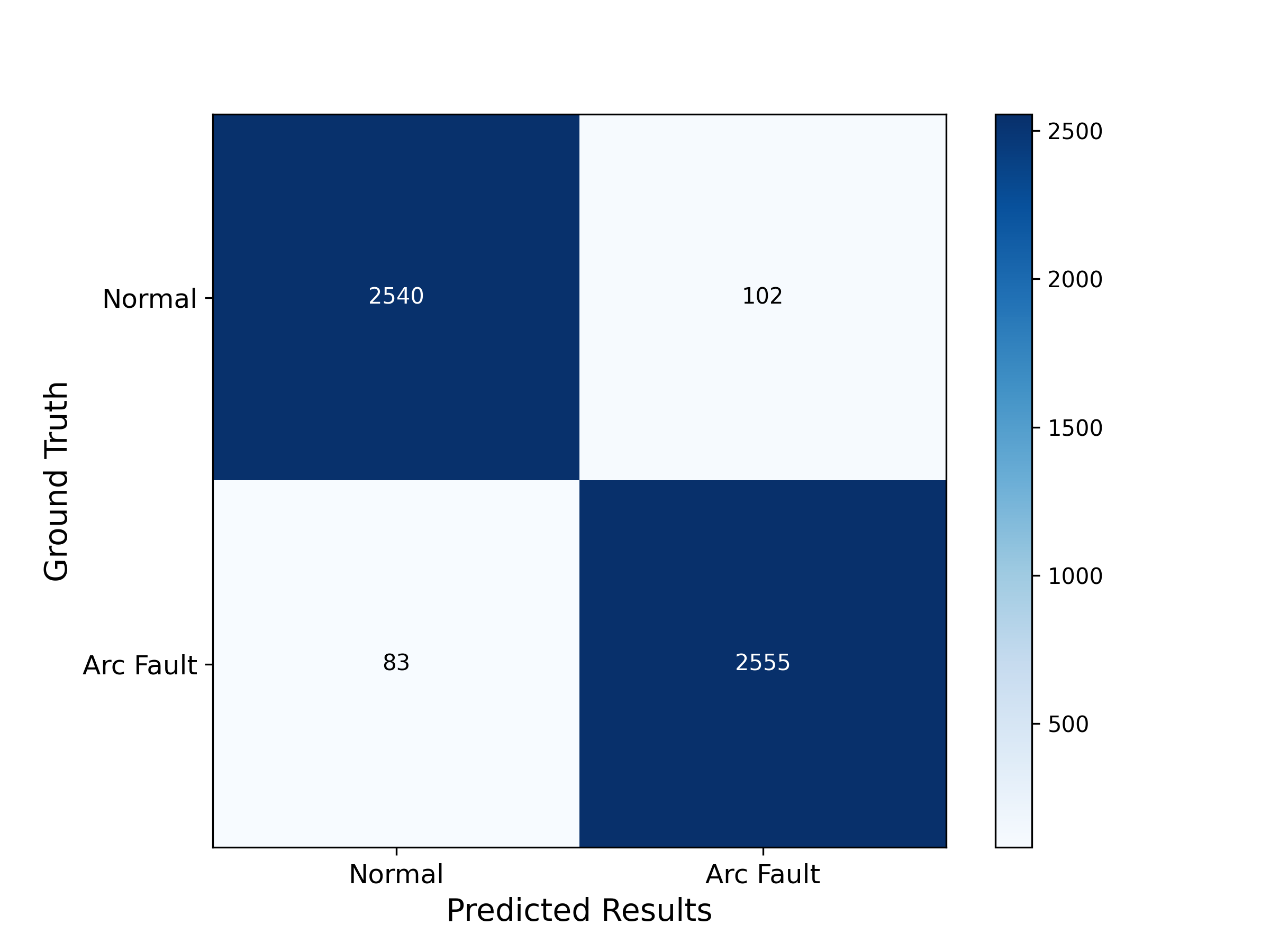}
    \caption{The confusion matrix of the classification model is used to test the classification accuracy of the model for two categories.}
    \label{fig:onfusion_matrix}
\end{figure}

\subsection{Visualization Experiment}
To visually demonstrate the training process of DCAMamba, we visualize the training results, as shown in Figure~\ref{fig:val_loss_acc}. The figure illustrates the dynamic changes in the validation loss and validation accuracy of the DCAMamba model during the training process. As the number of iterations increases, the validation loss (blue curve) rapidly decreases from a high value, stabilizes early on, and eventually converges to around 0.1. The validation accuracy (green curve) quickly rises, approaching 1, and remains stable. The results indicate that the model learns rapidly and achieves excellent performance during training, with no signs of overfitting throughout the process, further validating the effectiveness and stability of the model.

We also visualized the confusion matrix of the classification model, as shown in Figure~\ref{fig:onfusion_matrix}, which displays the classification results of the DCAMamba model on two types of samples: Normal and Arc Fault. Specifically, the number of samples correctly classified as Normal was 2540, and the number of samples correctly classified as Arc Fault was 2555. The number of samples that were falsely classified as Arc Fault when they were actually Normal was 102, while the number of samples falsely classified as Normal when they were actually Arc Fault was 83. The number of correctly classified samples was significantly higher than the number of misclassified samples, indicating that the model has a high classification accuracy and reliability.

\section{CONCLUSION}
In this paper, a novel arc fault detection model, DCAMamba, is proposed, which significantly enhances accuracy through the introduction of the Feature Amplification Strategy (FAS). The core idea of the FAS is to reorganize the fluctuating arc signals into signal representations that are easier for neural networks to learn. We have established a dedicated experimental platform, where various arc fault data were collected under different conditions to form a dataset. DCAMamba was systematically evaluated on this dataset. The experimental results show that the combination of FAS and DCAMamba can quickly and accurately identify arc faults, outperforming other methods with a significant speed advantage, and demonstrating strong industrial application potential.

\bibliographystyle{unsrt}
\bibliography{references}

\end{document}